# Hybrid plasmonic modes for enhanced refractive index sensing

*Bereket Dalga Dana[1,3], Ji Boyu[1], Jingquan Lin[1,4], Longnan Li[2], Alemayehu Nana Koya[2,5,]\*, Wei Li[2,]\**

[1]School of Physics, Changchun University of Science and Technology, Changchun 130022, China

[2]GPL Photonics Laboratory, State Key Laboratory of Luminescence and Applications, Changchun Institute of Optics, Fine Mechanics and Physics, Chinese Academy of Sciences, Changchun 130033, China

[3]Department of Physics, College of Natural and Computational Sciences, Jinka University, P. O. Box 165, Jinka, Ethiopia

[4]Zhongshan Institute of Changchun University of Science and Technology, Zhongshan 528400, China

[5]Department of Physics, College of Natural and Computational Sciences, Wolaita Sodo University, P. O. Box 138, Wolaita Sodo, Ethiopia

*E-mail: alemayehukoya@ciomp.ac.cn ; weili1@ciomp.ac.cn

Keywords: Localized surface plasmons, hybrid plasmons, charge transfer plasmons, Fano resonances, lattice plasmon resonance, refractive index sensing

Compared to single nanoparticles, strongly coupled plasmonic nanoparticles provide attractive advantages owing to their ability to exhibit multiple resonances with unique spectral features and higher local field intensity. These enhanced plasmonic properties of coupled metal nanoparticles have been used for various applications including realization of strong light-matter interaction, photocatalysis, and sensing applications. In this article, we review the basic physics of hybrid plasmonic modes in coupled metallic nanodimers and assess their potentials for refractive index sensing. In particular, we overview various modes of hybrid plasmons including bonding and antibonding modes in symmetric nanodimers, Fano resonances in asymmetric nanodimers, charge transfer plasmons in linked nanoparticle dimers, hybrid plasmon modes in nanoshells, and gap modes in particle-on-mirror configurations. Beyond the dimeric nanosystems, we also showcase the potentials of hybrid plasmonic modes in periodic nanoparticle arrays for sensing applications. Finally, based on the critical assessment of the recent researches on coupled plasmonic modes, the outlook on the future prospects of hybrid plasmon based refractometric sensing are discussed We believe that, given their tunable resonances and ultranarrow spectral signatures, coupled metal nanoparticles are expected to play key roles in developing precise plasmonic nanodevices with extreme sensitivity.

# 1. Introduction

Coupling metallic nanoparticles with small inter-particle separation results in interaction of elementary plasmons of the constituent nanoparticles and thus gives rise to a hybridized plasmon response[1,2].Compared to the single nanoparticle plasmon, the hybrid plasmonic mode has enhanced near-field intensity and tunable far-field responses with multiple resonances[3,4], implying its potentials for various plasmon-based applications including single-molecule spectroscopy[5,6], sensing[7], and optical trapping[8]. As a result, research on coupled plasmonic modes has grown exponentially with novel reports on various exotic modes of hybrid plasmons including bonding and antibonding dimer plasmons[9], charge transfer plasmons[10], gap plasmons[11], Fano resonances[12], and lattice plasmons[13].

On the other hand, optical sensing based on surface plasmon resonances has been extensively explored for several years. In principle, all types of plasmonic sensors take advantage of the collective oscillation of surface charges in metallic nanostructures, which can be broadly classified into propagating surface plasmon resonance (SPR) and localized surface plasmon resonance (LSPR). SPR-based sensors have been well-developed with commercialization of prototype sensors. However, since early 2000s[14], the research on LSPR-based sensing has been growing rapidly with ultimate goal of developing multifunctional and ultrasensitive sensors. Regardless of these efforts, the potentials of hybrid plasmonic modes for sensing application have not been properly assessed. In particular, given their tunable resonances and enhanced exotic optical properties, hybrid plasmonic modes can be applied for refractive index sensing.

To this end, in this review, we revisit the concept of plasmon hybridization with particular focus on the hybrid plasmon modes in dimeric nanostructures and assess their potentials for refractive index sensing. Specifically, based on our recent works [4,15-22] and other novel reports, we provide an overview of various modes of hybrid plasmons ranging from the simple dimer plasmon mode to the complex cavity mode. Furthermore, we also highlight research progresses on using the hybrid plasmonic modes for enhanced refractometric sensing. In this regard, we have covered bonding and antibonding modes in symmetric nanodimers, Fano resonances in asymmetric nanodimers, charge transfer plasmons in linked nanoparticles, hybrid plasmons in metallic nanoshells, gap plasmons in nanoparticle-on-mirror configurations, and lattice plasmon resonances in metallic nanoparticle arrays. The bulk sensitivity and figure of merit of each of the aforementioned hybrid plasmon modes are briefly discussed. Eventually, based on the assessment of the recent researches on coupled plasmonic modes, the outlook on the future prospects of hybrid plasmon based refractometric sensing are forwarded.

Finally, we would like to stress the fact that the scope of this review is limited to the hybridized plasmonic modes in coupled metallic nanoparticles that include anti-bonding and bonding dimer modes in symmetric nanoparticle dimers, Fano resonances (FRs) in asymmetric nanodimers, charge transfer plasmons (CTPs) in conductively coupled nanodimers, gap modes in nanoparticle-on-mirror (NPoM) geometries, and lattice plasmon resonance (LPR) in metallic arrays. Thus, it does not cover other hybrid plasmonic modes that can appear in complex plasmonic nanostructures like, for example, propagating surface plasmon polaritons (SPPs), hybridized LSPR-SPP modes, or Fabry–Pérot (F–P) cavity modes[23,24].

## 2. Refractive index sensing based on hybrid plasmonic modes

If two metallic nanoparticles are placed at distance approximately smaller than the size of the nanoparticles and if the coupled nanosystem is illuminated by incident light polarized along the interparticle axis, the interaction of the plasmonic properties of individual nanoparticles can be well described by the hybridization (PH) model[1,4,25,26]. The hybridized plasmonic modes in coupled metallic nanoparticles have been modified by controlling the inter-particle gap, nanoparticle morphology, and polarization of incident light [15-17,19,27]. Apart from observing enhanced optical responses, one can excite various types of hybrid modes including bonding and antibonding dimer modes in symmetric nanoparticle dimers[28], Fano resonances in asymmetric nanoparticle dimers[29], charge transfer plasmon modes in conductively linked plasmonic nanodimers[30], or gap modes in nanoparticle-on-mirror configurations[31].

These hybrid plasmonic modes show exotic properties including multiple resonances, narrow spectrum, ultrasensitive spectral shift, and hugely enhanced local field intensity − implying their potentials for sensing application[32,33]. One of the key parameters used to evaluate the performance of plasmonic sensors is sensitivity $S$[34], which is defined as change in SPR peak wavelength per change in refractive index. Quantitatively, it is expressed as $S = \Delta\lambda/\Delta n$, where $\Delta\lambda$ is change in resonance wavelength as a result of the change in dielectric environment $\Delta n$. Alternatively, to enable comparison between various modes that occur at different energies, a dimensionless quantity called Figure of Merit (FoM) is used. The FoM is generally defined as the ratio of the sensitivity $S$ and the SPR linewidth $\Gamma$ and is given by $FoM = S/\Gamma$[33]. In this section, we discuss various exotic modes that can arise in coupled plasmonic nanoparticle dimers of various configurations and assess their potentials for refractive index sensing.

### 2.1. Bonding and anti-bonding modes in symmetric nanodimers

In simple plasmonic nanodimers, like strongly coupled metallic nanoparticle dimers, the interactions between the plasmon resonances of constituent nanoparticles results in splitting of the hybrid resonance into two new modes (see **Figure 1A&B**): the lower energy bonding dimer modes (BDMs) and the higher energy antibonding dimer modes (ABDMs)[1]. These two modes differ in their coupling to incident light that the bonding dimer modes have mutually aligned longitudinal dipoles, resulting in plasmonic modes that are electric dipolar in nature and readily couple to the far field, thus called bright modes[2,35]. Therefore, the optical responses of plasmonic nanodimers are dominated by bonding dimer modes, as the BDMs are highly dependent on the polarization of the incident light, illumination angle, dimer geometry as well as dimer gap, and dielectric environment. In particular, for polarization of the incident light along the dimer axis, the hybridized bonding dimer plasmon resonance tends to redshift as the dimer gap decreases, with increase in intensity. However, when the light is polarized orthogonal to the dimer axis, the plasmonicc coupling is weak, thus a small blue shift of the individual nanoparticle plasmon is observed [16,36-38]. On the other hand, in the higher energy collective dipolar mode, known as the anti-bonding dimer mode, the dipoles on each individual nanoparticle are anti-aligned, resulting in no net dipole moment and inability of this mode to couple to the far field, thus the ABDM is called dark or cavity-like mode[2]. This type of hybrid plasmonic mode is generated due to the effect of symmetry breaking in nanoparticle dimers[35]. Generally, the resonance peaks and intensities of both the antibonding and bonding dimer modes sensitively depend on the orientation of nanodimers and their dielectric environment, implying their potentials for sensing application.

In this regard, Beuwer and Zijlstra experimentally and numerically demonstrated that gold nanorod dimers have both bonding and antibonding plasmon modes in the visible to near-infrared wavelength regime[33]. For nanorod dimers, The scattering spectra displayed in Figure 1C&D clearly show the antibonding (higher energy) and bonding modes (lower energy) that arise due to hybridization of the longitudinal SPRs of the individual nanorods. The angle and spacing between the nanorods determine the wavelength and relative intensity of the two peaks. To assess the refractometric sensing performance of individual nanorod dimers, they measured the scattering spectra of dimers in three different media: water (n = 1.33), a mixture of water and ethylene glycol (n = 1.38), and ethylene glycol (n = 1.43). For both configurations of the dimers, redshift is observed when increasing the refractive index of the environment. They also made comparison of the sensitivity $S$, resonance linewidth $\Gamma$, and *FoM* for both bonding and antibonding modes as a function of dimer angle. It was found that, in presence of coating of the nanodimers, the antibonding mode is barely affected whereas the bonding mode exhibits a significant reduction in its sensitivity (see Figure 1E&F).

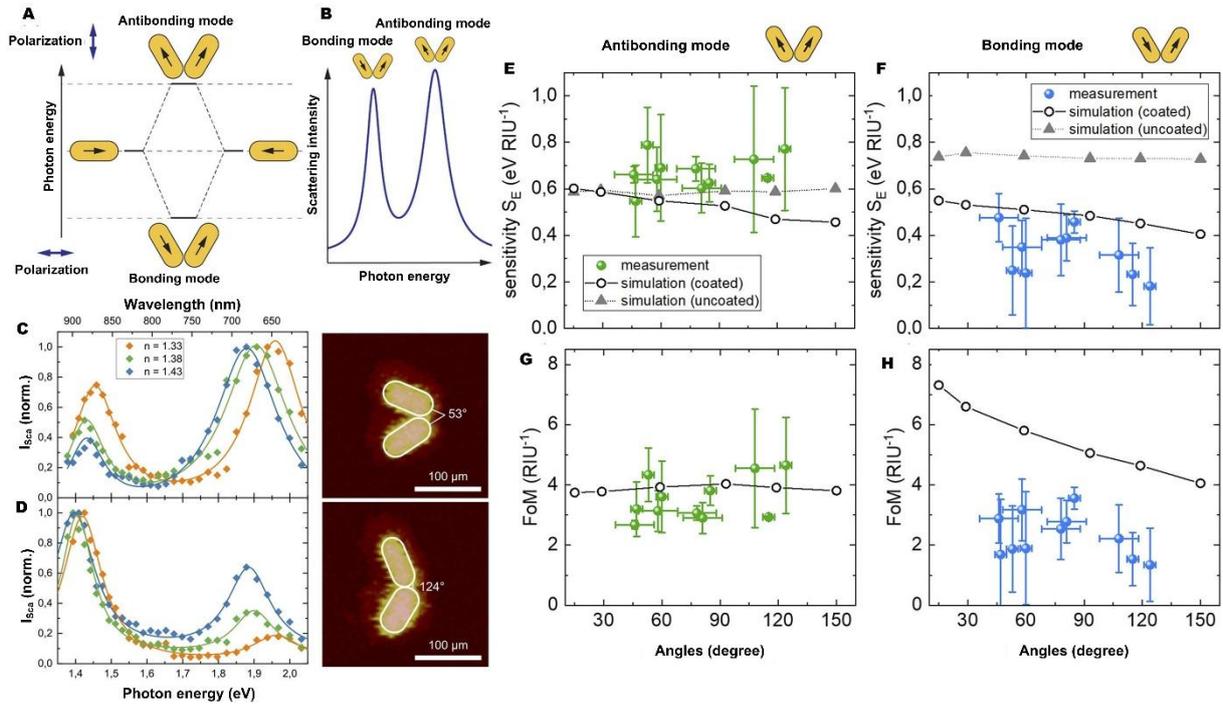

**Figure 1.** Refractometric sensing based on bonding and antibonding plasmonic modes in gold nanorod dimers. A) Hybridization of the longitudinal SPR in nanorod dimers. Depending on the polarization of the excitation, the antibonding or bonding mode is excited. B) The resulting scattering spectra exhibit two resonances whose wavelength and relative intensity depend on the relative particle size, spacing, and orientation. C) and D) Measured dark-field scattering spectra of nanorod dimers as a function of change in refractive index for the interparticle orientation angles of 53 °(top) and 124 °(bottom). These spectra clearly show the antibonding (higher energy) and bonding modes (lower energy) that arise due to hybridization of the longitudinal SPRs of the individual particles. The right column shows corresponding AFM images of the same dimers. E) and F) Refractive index sensitivities of antibonding and bonding modes, respectively, as a function of dimer angle. The open circles (solid triangles) indicate simulations for a coated (uncoated) dimer consisting of two particles of 72 nm in length and 30 nm in diameter. G) and H) Figure of merit (FoM) of the antibonding and bonding modes of the same particles. Reproduced with permission from[33]. Copyright 2021 Authors. Published under an exclusive license by AIP Publishing.

They also demonstrated that the antibonding mode has a broader linewidth due to the proximity of the interband transitions of bulk gold at energies larger than 2 eV whereas the narrower bonding mode exhibits the opposite trend because of the radiation damping effect. The FoM for the antibonding mode is nearly independent of the dimer angle and for the bonding mode on the other hand, the FoM is dictated by an interplay between a decreasing sensitivity combined with an increasing linewidth as a function of dimer angle [see Figure 1G&H].

## 2.2. Fano resonances in asymmetric nanoparticle dimers

On the other hand, breaking the compositional or geometric symmetry of plasmonic nanosystems gives rise to an asymmetric resonance mode called Fano resonance. However, only few nanostructure configurations can support deep Fano resonance with strong polarization dependence. Since the relevant plasmonic interaction can be tuned by changing the morphology and composition of nanostructures, plasmonic nanodimers have confirmed to be an ideal choices to produce FR modes with sharp dispersion[39]. In this regard, Qin et al. experimentally and numerically demonstrated the observation of deep Fano resonance with strong polarization dependence in Au nanoplate–nanosphere heterodimers[40]. The microscopic origin of these modes in asymmetric dimers arises from the destructive interference between super-radiant and subradiant plasmon modes (see Figure **2A – E**). Generally, the key elements underlying these modes are radiative interference and the effect of symmetry breaking[2,41]. As it was numerically demonstrated that, for symmetric nanodimers, no FR is observed whereas asymmetric dimers can lead to the manifestation of these modes due to broken symmetry[19]. Thus, introducing either morphological or compositional asymmetry into plasmonic nanodimers is the common way to produce deep Fano resonances and to manipulate their spectral features. In addition, simultaneous introduction of both the compositional and geometrical asymmetries to plasmonic nanodimers can induce double Fano resonances[42,43].

Given its narrow line width, sharp asymmetric spectral shape, low radiation loss, huge modification ability, and ultrasensitivity to small alteration in the dielectric environment, the potential of Fano resonance for refractive index sensing is paramount[44]. In this regard, Dana *et al.* have explored the bulk sensitivity of FR in nanoring-nanodisk asymmetric gold dimers[21]. The refractive index sensing performance of the FR is determined in terms of its quality factor and figure of merit as a function of the geometric parameters of asymmetric nanodimer and polarization angle of the incident light. They reported comparatively higher quality factor (about 18.5) for Fano resonances in asymmetric disk-ring nanodimer. Moreover, as it was numerically demonstrated, the Q factor exponentially decreases as the size of the nanodisk increases while the relative resonance intensity of FR increases at the same time. Moreover, to enhance sensing applications in the shape-asymmetric dimer, they also studied the potentials of Fano resonances for refractive index sensing and reported about ~ 560nm/RIU at optimum geometry and polarization (see Figure 2F & G). Thus, the problem of low quality factor in plasmon that has hindered its applications can be overcome by designing plasmonic nanosystems sustaining hybrid resonances with high quality factors. To this end, by breaking the nanofins' out-of-plane symmetry in parameter space, Aigner *et al.* most recently achieved high-quality factor (up to 180) modes under normal incidence[45]. Thus, to enhance the performance of hybrid plasmon sesnors, it is important to design nanodimers with asymmetric configurations with lower radiative losses and high quality factors[46].

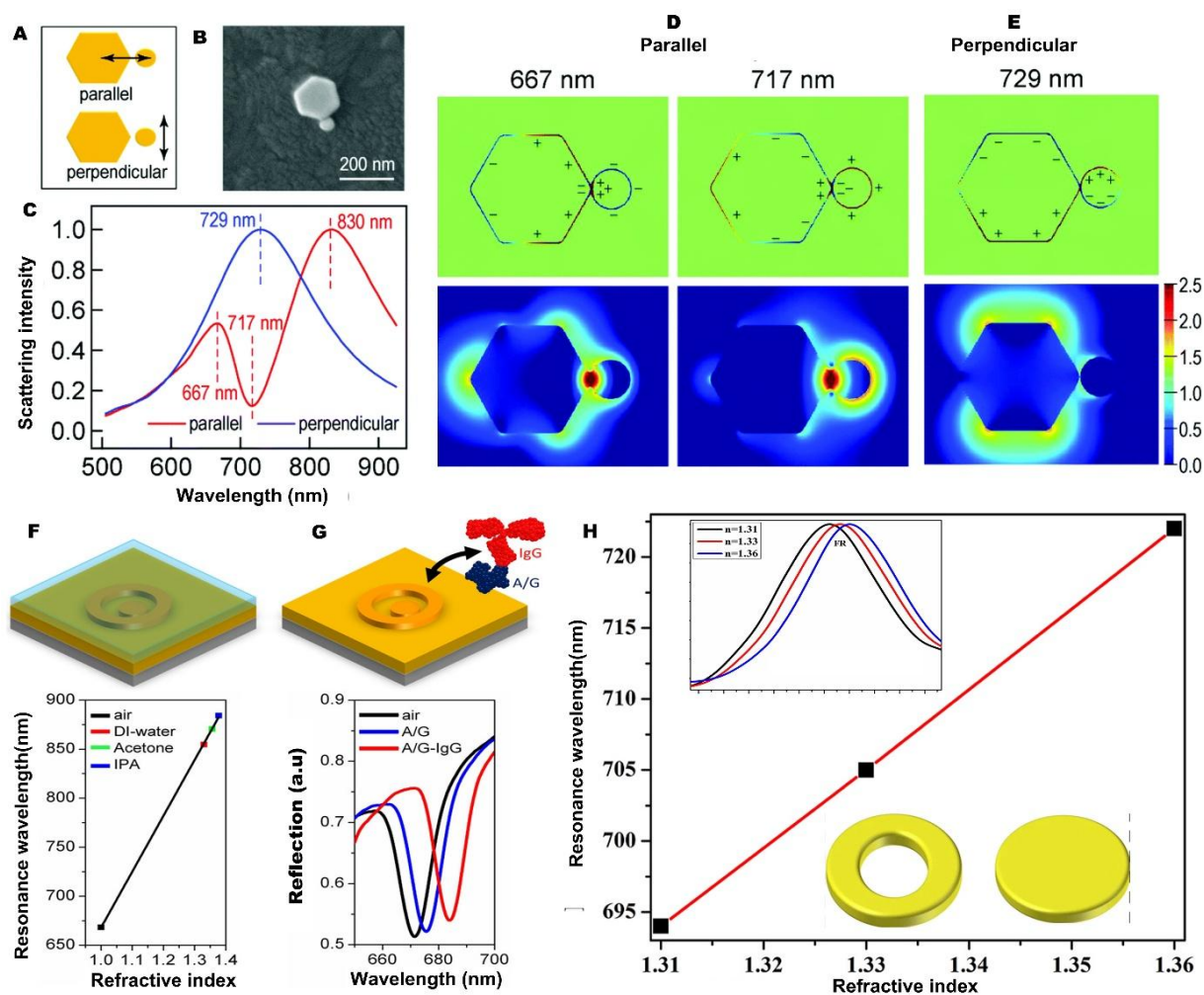

**Figure 2**. Fano resonances in asymmetric plasmonic hetrodimers and their implications for refractive index sensing. A) Schematic of in-plane parallel and perpendicular excitation polarization states relative to the Au nanoplate (NPL)–nanosphere (NS) heterodimers. B) Typical SEM image of the NPL-NS dimer. C) Scattering spectra under the in-plane parallel and perpendicular excitations. D) and E) Corresponding charge distributions (top row) and near-field profiles (bottom row), respectively, for parallel and perpendicular excitations. Reproduced with permission from[40]. Copyright 2017, The Royal Society of Chemistry. Fano resonance spectral shifts as a result of F) change in refractive index and G) introducing protein (A/G) and antibody (IgG). Reproduced with permission from[29]. Copyright 2012, American Chemical Society. H) Refractive index sensing with FRs in Au nanoring-nanodisk asymmetric dimers. Reproduced with permission from[21]. Copyright 2022, Elsevier B.V.

## 2.3. Charge transfer plasmons in linked nanoparticle dimers

Conductive linking of metallic nanodimers gives rise to excitation of a novel plasmonic mode called charge transfer plasmon resonance[47]. CTP modes have also been witnessed in sub-nanometer capacitive gaps of plasmonic dimers, where a direct transfer of charges takes place by quantum tunneling effect[48]. Apart from the lower energy CTP mode, one can also observe other modes like screened bonding dimer plasmon (SBDP) mode and even Fano resonances in conductively linked plasmonic dimers with broken symmetry (see Figure **3A&B**)[22,49]. The SBDP mode is identified as a blueshifted bonding mode owing to a weaker polarization of the charge distribution of the individual nanoparticles and the CTP mode is attributed to the electron density oscillates between the two particles, making one nanoparticle momentarily positively and the other negatively charged[50]. Nevertheless, the CTP modes have striking spectral features including ultratunable mid-infrared spectrum that can be easily tuned by controlling the conductance of bridging nanowire and geometries of linked nanoparticles (see Figure 3C & D)[15,51]. Thus, understanding the direct transfer of charges in conductively bridged plasmonic nanodimers is essential for developing plasmon-based nanodevices including nanomotors, sensors and optoelectronic devices[10,18,52].

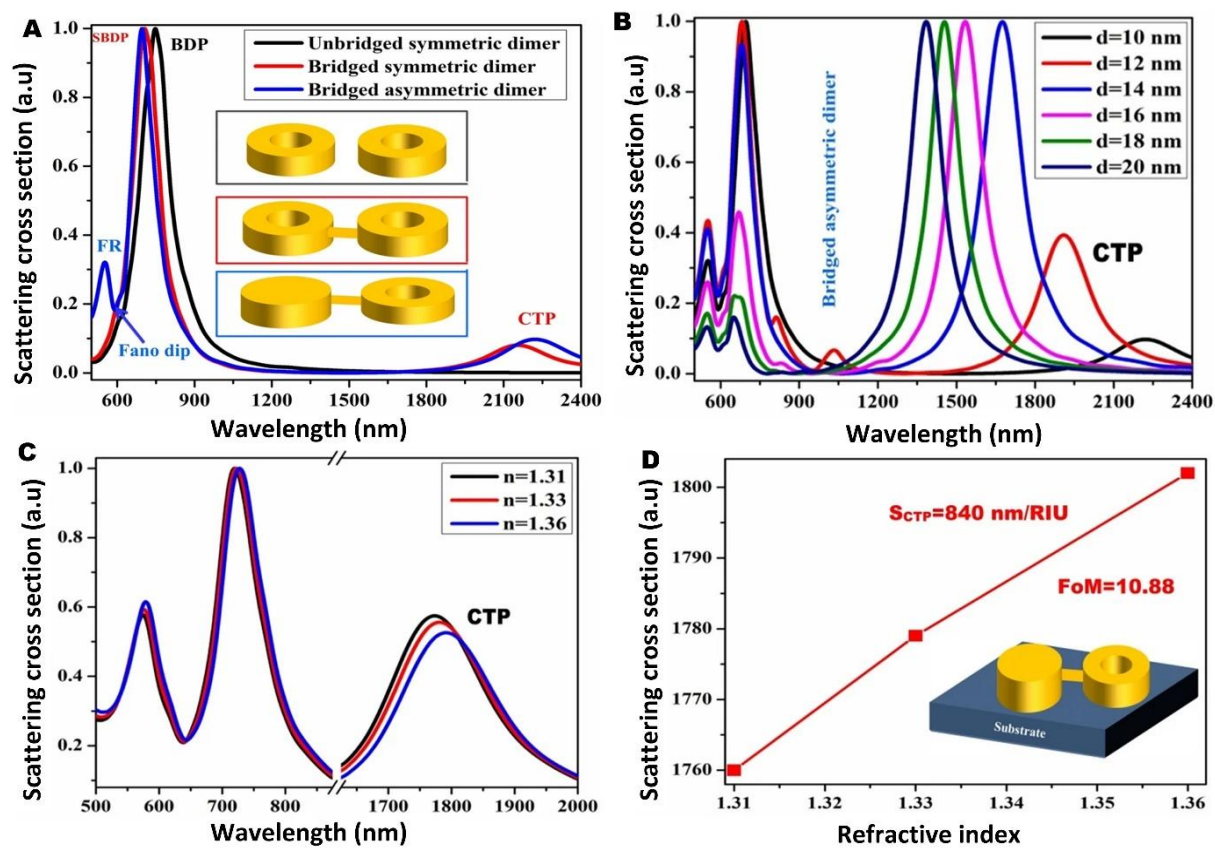

**Figure 3**. Charge transfer plasmon resonance based on refractive index sensing. (a) Scattering spectral features of bridged symmetric and asymmetric nanoparticle dimers. The bonding dimer plasmon (BDP) spectrum of unbridged linked nanodimer is presented for comparison. When the geometric symmetry of the linked nanodimer is broken, the scattering spectra exhibits several resonance peaks including CTP mode at longer wavelength, screened bonding dimer plasmon (SBDP) mode, and Fano resonance. B) Tuning the CTP resonances in bridged asymmetric nanodimer by controlling junction diameter d. C) and D) Sensitivity of bridged asymmetric nanodimer as a function of the refractive index of the substrate. Reproduced with permission from[22]. Copyright 2022, The Authors, under exclusive licence to Springer-Verlag GmbH, DE part of Springer Nature.

In particular, given its ultratunable and narrow spectral signatures, the CTP resonance has great potential for bulk, surface, gas and molecular sensing[53]. In this regard, Dana *et al.* have numerically explored the sensing abilities of bridged symmetric and asymmetric plasmonic nanodimers for refractive index sensing application[22]. It was found that the refractive index sensitivities of CTP modes ($S_{CTP}$) bridged symmetric and asymmetric dimers are approximately 540nm/RIU and 840nm/RIU, respectively (see Figure 3E & F). This indicates that the bridged asymmetric dimer has better sensitivity than the bridged symmetric dimer. Moreover, the values of FoM calculated for the CTP modes of bridged symmetric and asymmetric dimers are also found to be higher, 6.03 and 10.88, respectively. Compared to the sensitivity and FOM obtained from Fano resonance[19,21] the CTP modes have higher values, implying that the CTP modes in bridged nanoparticle dimers are the best candidate for normal LSPR based sensing[18,54].

## 2.4. Hybrid gap modes in particle-on-film plasmonic nanosystems

Gap modes sustained by nanoparticle-on-mirror configurations have attracted extensive research interest due to flexible control over their spectral response and significantly enhanced field intensities at the particle–film junction[55]. Le *et al.* demonstrated that, polarization-controlled excitation of NPoM dimer made of a 200 nm diameter Au sphere placed 2 nm above 45 nm thick Au film can feature three different resonances in the scattering spectra (see **Figure 4A-F**), which originate from tightly confined gap modes having different azimuthal characteristics[11]. These hybrid modes with extremely small mode volume and strong local field intensity have been attractive for exploring novel light-matter interaction phenomena and plasmon-enhanced spectroscopy[31]. Wang *et al.* extended these efforts by designing and experimentally demonstrating plasmonic nanocavity made of a hexagonal Au nanoplate positioned over an ultrathin Au film[56].

These hexagonal Au NPoM nanostructures sustain excitation of plasmonic radial breathing modes (RBMs), which arise from the surface plasmon waves confined in the flat nanoparticles. The RBMs have an extremely low radiation yet a remarkable intense local field, implying their potentials for surface-enhanced Raman spectroscopy and sensing. To this end, Wang and co-workers developed a plasmonic nanocavity sensor consisting of a hexagonal Au nanoplate positioned over an ultrasmooth Au film[57] (see Figure 4G). Compared with refractive index sensors based on the localized surface plasmon resonance, refractive index sensors based on the plasmonic radial breathing modes have strongly reduced radiative damping, yielding an outstanding FOM of 11.2 RIU$^{-1}$ (see Figure 4H). Therefore, the hybrid plasmonic modes arising from NPoM configuration may open up a new route in exploring such modes for biological and chemical sensing.

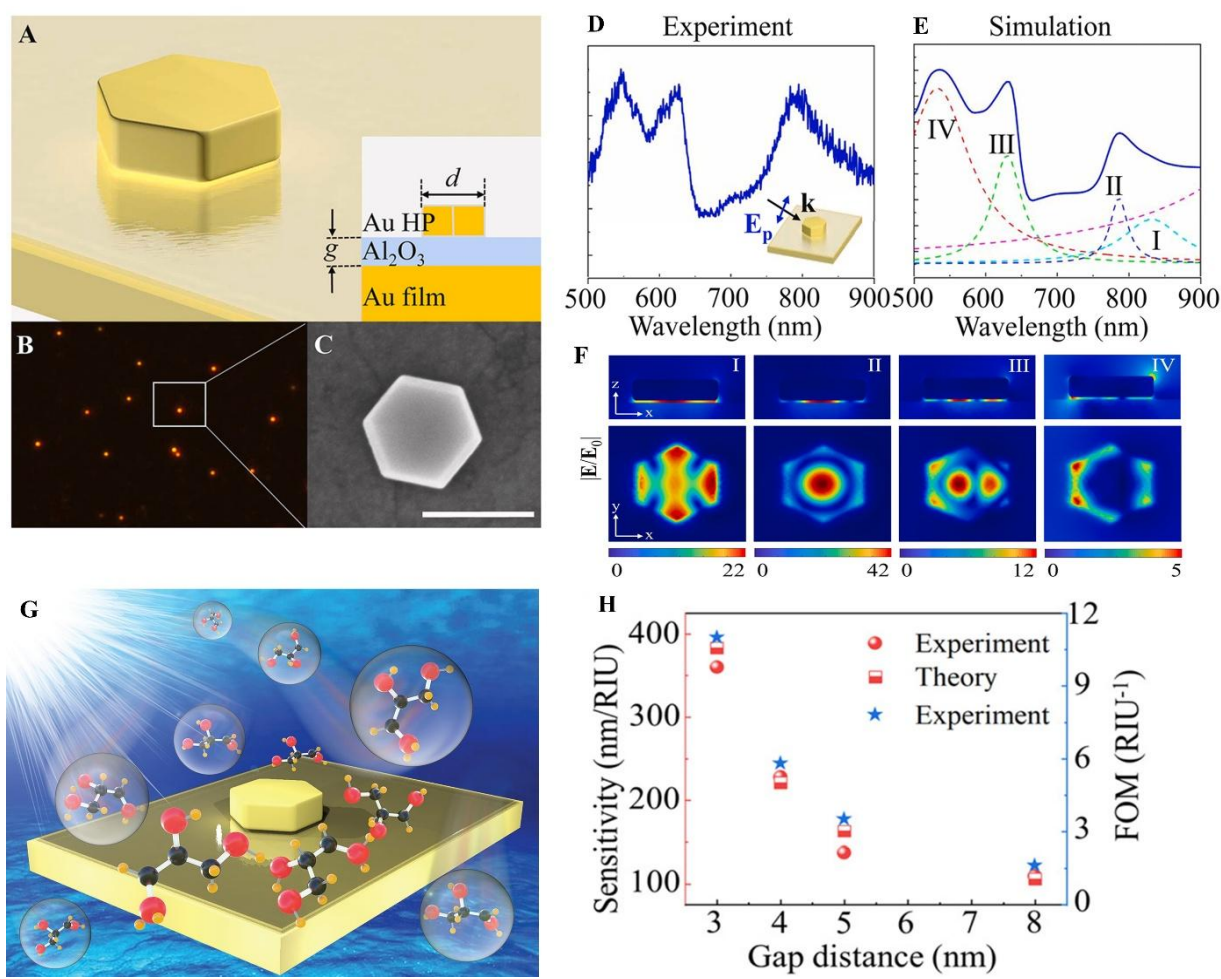

**Figure 4**. Sensing characteristics of hybrid gap modes that arise in nanoparticle-on-mirror (NPoM) configuration of plasmonic nanostructures. A) Schematic of Au NPoM geometry consisting a hexagonal nanoplate, an ultrasmooth Au film and an Al2O3 spacer. Inset shows the cross-section diagram of hexagonal Au NPoM. B) Dark-field image and C) SEM image of the NPoM. D) Experimental and E) simulated scattering spectra of the hexagonal Au NPoM (d = 185 nm, g = 5 nm) illuminated by p-polarized beams. F) Corresponding Electric field enhancements of modes I–IV with spectral positions labeled in E). Reproduced from[56] under the Creative Commons Attribution 4.0 International License. Copyright 2021, Qifa Wang et al., published by De Gruyte. G) Schematic of refractive index sensor based on nanoplate-on-mirror nanocavity. H) The dependence of refractive index sensitivity and FOM on nanocavity gap size. Reproduced with permission from.[55] Copyright 2022, The Royal Society of Chemistry.

## 2.5. Coupled plasmonic modes in nanoshell dimers

Metallic nanoshells have versatile geometries that give freedom to design plasmonic nanostructures with enhanced and tunable optical responses. The early works in this field by Nordlander and co-workers demonstrated that as two nanoshells are brought closer to each other, dimer plasmons are formed by hybridization of individual nanoshell plasmons[58]. The plasmonic responses of nanoshells can be further tailored by breaking their geometric symmetries[26,59]. Nevertheless, as of the plasmonic coupling in simple nanoparticle pairs, the strength of plasmonic coupling in complex nanoshell geometries, like dielectric core-metal shell nanostructure[36] or metal−dielectric−metal multilayer nanoshells[60] decreases as the interparticle gap increases. This universal scaling behavior of coupled plasmonic nanostructures have been utilized for designing ultrasensitive LSPR-based plasmonic sensors.

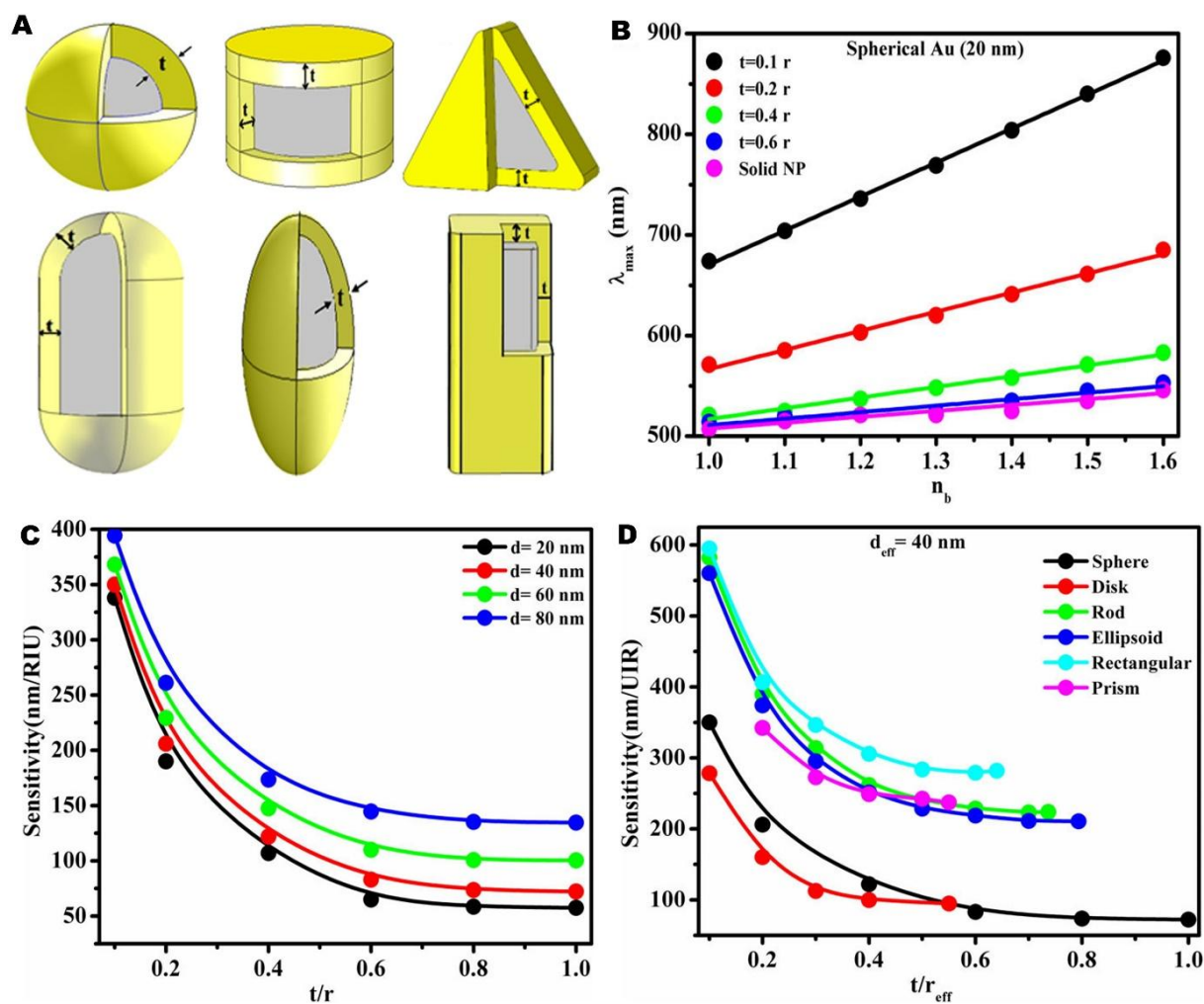

**Figure 5**. Hollow-gold nanoshells for refractive index sensing. A) Schematic of spherical, disk, triangular prism, rod, ellipsoid, and rectangular block hollow/Au nanoshells. B) LSPR peak wavelength shift of the spherical hollow/Au nanoshell (with the diameter of d = 20 nm) as a function of change in refractive index of the surrounding medium. C) Size effect on the LSPR sensitivity of the spherical hollow/Au nanoshells for the NPs with the total diameter of d = 20 nm, d = 40 nm, d = 60 nm, and d = 80 nm. D) Influence of shape and shell thickness on the LSPR sensitivity of the sphere, disk, rod, ellipsoid, rectangular block, and prism shaped hollow/Au nanoshell with an effective diameter of $d_{eff}$ = 40 nm. Lines are added as guides to the eye. Reproduced with permission from[61]. Copyright 2019, Authors. Published under license by AIP Publishing.

In this regard, Shabaninezhad and Ramakrishna recently studied the refractive index sensitivity of hollow gold nanoshells using theoretical modeling where the influences of shape, size, shell thickness, and aspect ratio are addressed[61]. They investigated various shapes of hollow Au nanoshells including sphere, disk, triangular prism, rod, ellipsoid, and rectangular block (see **Figure 5A**) to determine the LSPR peak position and LSPR sensitivity. As displayed in Figure 5B, the LSPR sensitivity increases with reducing shell thickness at an exponential rate. As an example, for NP with d = 20 nm, the LSPR sensitivity increases significantly from 57.5 to 338 nm/RIU by reducing shell thickness from t = 1 r (solid sphere) to t = 0.1r, where r is radius. Moreover, as displayed in Figure 5C, the sensitivity of LSPR significantly increase with reducing ratio of shell thickness-to-total radius (t/r), which is attributed to the enhanced inner and outer layers' plasmonic coupling due to reducing the shell thickness. It was theoretically shown that the rectangular block and rod-shaped Au nanoshells have maximum LSPR sensitivity when compared to other shaped Au nanoshells (see Figure 5D). In addition, increased sensitivity was observed for higher aspect ratio as well as for smaller shell thicknesses, which are characteristic effects of plasmon hybridization.

### 3. Plasmonic sensors based on nanoparticle arrays

A periodic array of metallic nanoparticle assemblies can exhibit exciting new collective properties which are different from those of individual nanoparticles or corresponding bulk materials[62]. As a result, ordered arrays of metal nanoparticles offer new opportunities to engineer light-matter interactions through the hybridization of localized surface plasmon resonances of the constituent nanoparticles. The generated lattice plasmon resonances exhibit much higher quality factors compared to those observed in isolated metal nanoparticles (see **Figure 6A-C**)[13,63]. Apart from the enhanced far-field responses, the near-field properties of metallic nanoparticle arrays also show interesting phenomena. As a result of plasmon hybridization effect, the assembly of plasmonic nanoparticles generate extremely high local fields (hot spots) confined at the interparticle gaps[64], which have been widely explored as substrates for surface-enhanced Raman spectroscopy (SERS)[65].

The optical properties of plasmonic nanoparticle arrays are often investigated by varying the lattice parameter as well as the size and shape of the building blocks (see Figure 6D-E), which enable control over both near-field and far-field plasmonic coupling[66,67]. The fine tuning of plasmonic coupling in metallic superlattices has shown promising implications for sensing, catalysis, and plasmonic heating[68]. In particular, the ultranarrow spectral features and thus high quality factor of the LPR are interesting for such applications as nanolasing and refractive index sensing[69]. This evident from the abrupt and sensitive transmission spectral shift as a result of change in the dielectric environment of Au nanoparticle arrays (see Figure 6F-I). Regardless of these facts, however, the potentials of LPR of metallic nanoarrays for sensing application yet to be exploited.

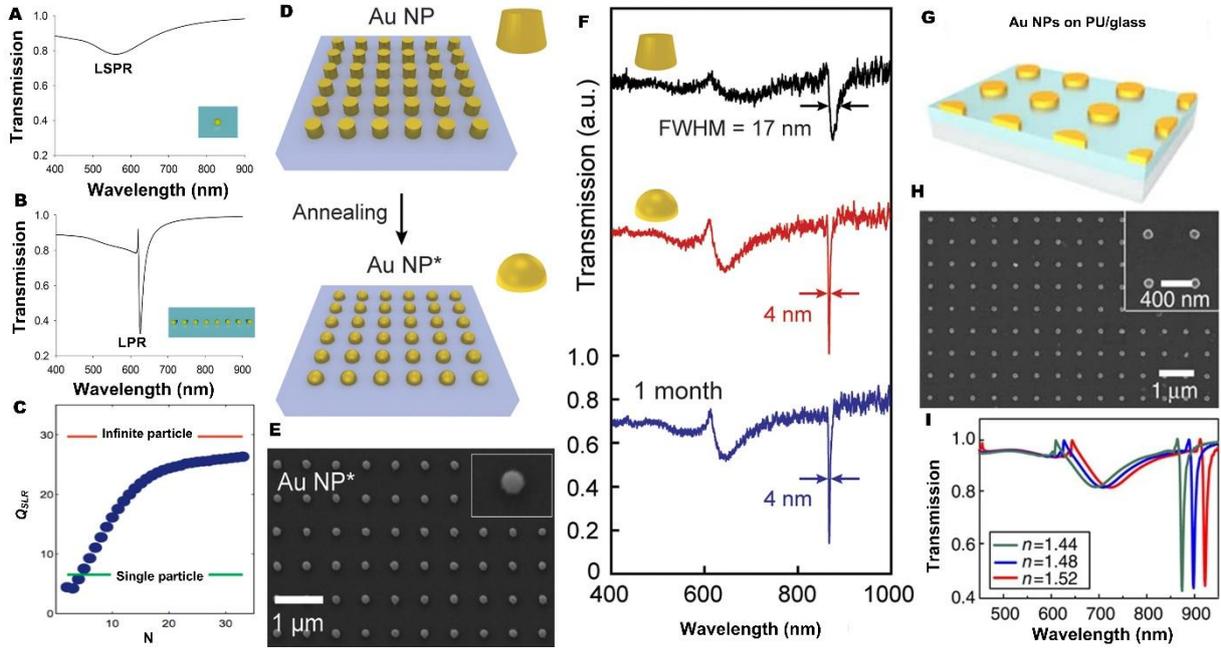

**Figure 6**. The potentials of lattice plasmon resonances (LPRs) of metallic nanoparticle arrays for refractive index sensing. Transmission spectra showing striking difference between A) localized surface plasmon resonance of single nanoparticle and B) LPR of nanoparticle arrays. Reproduced with permission from[66]. Copyright 2018, American Chemical Society. C) Q-factor of the lattice plasmon resonance ($Q_{LPR}$) for arrays of gold nanodisks (with diameter of 120 nm size and 50 nm height) as a function of the number of particles N along the side of the square array. Reproduced with permission from[70]. Copyright 2012, Elsevier B.V. D) Scheme of annealing process for Au nanoparticle (Au NP) lattice. E) Scanning electron microscopy (SEM) images of Au NP arrays after annealing process. F) Measured transmission spectra of untreated Au NP lattices (black) and Au NP* lattices at 1 d (red) and 1 mo (blue) after annealing treatment. Reproduced from[71]. under a Creative Commons Attribution 4.0 International License. Copyright 2020, Published under the PNAS license. G) Schematics of Au NP arrays on polyurethane (PU) on glass (PU/glass) substrate. H) Scanning electron microscope image of Au NPs on PU/glass. I) Transmission spectra of Au NPs in different dielectric environments (n=1.44, 1.48 and 1.52). Reproduced from[69]. under a Creative Commons Attribution 4.0 International License. Copyright 2015, Macmillan Publishers Limited.

## 4.    Conclusions

In this review, we have overviewed the spectral characteristics of hybrid plasmons that emerge in coupled metallic nanodimers and also assessed the potentials of such modes for refractive index sensing. In particular, we have reviewed the fundamental physics of hybrid plasmons in strongly coupled metal nanodimers and depending on the polarization of the excitation, both antibonding and bonding modes can be excited. The resulting scattering spectra exhibit two resonances whose wavelength and relative intensity strongly depend on the nanoparticle size, spacing, and orientation. The presence of two narrow plasmon resonances in coupled metallic nanoparticles especially in nanorod dimers makes them exciting candidates for multicolor and multiplexed sensing.

On the other hand, if the geometric or compositional symmetry of the plasmonic nanodimer is broken, one can observe Fano-type resonances in the hybrid plasmon resonance, which is ascribed to the hybridization of bright and dark modes. Comparatively, Fano resonances have high quality factor indicating that they can be exploited for sensing applications. We have also overviewed recent developments in charge transfer plasmon resonances that arise in conductively linked plasmonic nanodimers. CTP resonances are extremely sensitive to the geometric parameters of the nanolinkers and thus can be good candidates for enhanced refractive index sensing. Furthermore, by coupling plasmonic nanoparticles with metallic thin films, one can excite hybrid gap plasmon modes that are confined in the cavity of the nanoparticle and film, suggesting their potentials for realizing ultrasensitive sensors that can detect objects with atomic sizes. Similarly, the hybrid plasmonic modes sustained by metallic nanoshells have tunable resonances that strongly depend on the nanoshell thickness. Finally, we have also assessed the sensing capability of plasmonic nanoparticle arrays that sustain ultranarrow spectral features.

The aforementioned and other hybrid plasmonic modes in coupled metallic nanoparticles can demonstrate multiple resonances with ultranarrow spectral features and hugely enhanced local field intensities, implying their potentials for enhanced refractive index sensing. Particularly, the ultratunable resonance of CTP modes and ultranarrow spectral features of lattice plasmon resonances can play key roles in developing ultrasensitive refractive index sensors. Recent theoretical work on plasmonic mode analysis in bowtie-shaped Au nanoantenna also demonstrated that the complex refractive index, $n + ik$, of the substrate can provide additional design flexibility in tailoring the spectral characteristics of the antenna mode as well as the spatial distribution and attainable field enhancements of its associated near-field hotspots [72]. This implies that plasmonic nanodimers are promising platforms for next-generation spectroscopy, sensing and ultrafast information processing technologies [73,74].Thus, further research on coupled plasmonic nanoparticles with actively tunable optical responses and precisely controlled geometries[75,76] is crucial for the development of precision plasmonics and plasmonic nanodevices.

**Acknowledgements**

This work was funded by National Natural Science Foundation of China (Grant No. 62134009, 62121005); and Chinese Academy of Sciences President's International Fellowship Initiative (Grant No. 2023VMC0020). B. D. D and A. N. K contributed equally to this work.